# An Effective Deep Learning Based Multi-Class Classification of DoS and DDoS Attack Detection

**Arun Kumar Silivery**
Department of Computer Science & Engineering,
University College of Engineering, Osmania University, Hyderabad- 500 007
arunsilivery@osmania.ac.in

**Kovvur Ram Mohan Rao**
Department of Information Technology
Vasavi College of Engineering, Osmania University, Hyderabad- 500 031
krmrao@staff.vce.ac.in

**L. K. Suresh Kumar**
Department of Computer Science & Engineering,
University College of Engineering, Osmania University, Hyderabad- 500 007
suresh.l@uceou.edu

***Abstract*** *– In the past few years, cybersecurity is becoming very important due to the rise in internet users. The internet attacks such as Denial of service (DoS) and Distributed Denial of Service (DDoS) attacks severely harm a website or server and make them unavailable to other users. Network Monitoring and control systems have found it challenging to identify the many classes of DoS and DDoS attacks since each operates uniquely. Hence a powerful technique is required for attack detection. Traditional machine learning techniques are inefficient in handling extensive network data and cannot extract high-level features for attack detection. Therefore, an effective deep learning-based intrusion detection system is developed in this paper for DoS and DDoS attack classification. This model includes various phases and starts with the Deep Convolutional Generative Adversarial Networks (DCGAN) based technique to address the class imbalance issue in the dataset. Then a deep learning algorithm based on ResNet-50 extracts the critical features for each class in the dataset. After that, an optimized AlexNet-based classifier is implemented for detecting the attacks separately, and the essential parameters of the classifier are optimized using the Atom search optimization algorithm. The proposed approach was evaluated on benchmark datasets, CCIDS2019 and UNSW-NB15, using key classification metrics and achieved 99.37% accuracy for the UNSW-NB15 dataset and 99.33% for the CICIDS2019 dataset. The investigational results demonstrate that the suggested approach performs superior to other competitive techniques in identifying DoS and DDoS attacks.*

## 1. INTRODUCTION

In today's world, systems for Information and Communication Technology (ICT) significantly influence every element of society and business. At the same time, cyberattacks on ICT systems are becoming more sophisticated and more frequent [1-3]. This significantly impacts network performance, resulting in instability that would render the network unusable. Thus, ICT systems require a very effective network security solution. One of the most popular methods for spotting different kinds of malicious network attacks is Intrusion Detection System (IDS).

The two primary strategies for detecting intrusion are signatures-based IDS and anomalies-based IDS [4-6]. The signature-based IDS is also known as Knowledge-based Detection or Misuse Detection. It is as effective as upgrading the database at a specified time since it focuses on finding a "signature" or patterns of intrusion event. The anomaly-based IDS is also known as Behavior-based Detection. It is based on comparing reliable behavioral patterns with unexpected behaviors while observing routine activities [7-9]. An administrator uses an Intrusion Prevention System (IPS) to stop threats like denial of service (DoS), Distributed DoS (DDoS) attacks, Trojan horses, etc., when the IDS system detects them.

The DoS and DDoS attacks are significant traits in anomaly-based IDS. As a result of the modification of various services, there has been a growth in these attacks over the past ten years, establishing them as a severe threat to the security of networks [10-12]. Additionally, tracking security attacks has become a significant hurdle for most organizations despite spending excessive money to secure the system. However, large-scale cyberattacks are still happening, attackers are becoming more sophisticated, and tools to defend against them have become outdated.

Every DDoS assault is based on the same basic idea. Using a network protocol, the attacker floods the server with spoof request packets. Since the target server cannot distinguish between them, it begins serving every packet. The server becomes overloaded and crashes while attempting to fulfill all the requests [13-15]. Because of this, an attacker can exhaust all server resources, resulting in a denial of service. It is referred to as a distributed denial of service attack. In this, multiple devices send packets instead of a single source machine.

In the past few years, many IDS techniques have been presented based on various approaches, such as mathematical formulations and data mining techniques like machine learning. Poor performances are caused by the difficulty in managing the high-dimensional network traffic data by these statistical formulations and conventional machine learning models [16-18]. Furthermore, most existing techniques used only binary classification, such as whether it is an attack. Therefore, better approaches are required for IDS, such as deep learning-based techniques. Due to its powerful learning and feature extraction capabilities, particularly in scenarios involving large datasets, deep learning has been widely recommended for IDS in recent years. Multiple layers are used in deep learning approaches to gradually extract essential features from the raw input without domain knowledge [19, 20].

Therefore, an Alexnet-Resnet-50-based deep learning model is presented in this paper for multi-class classification of DoS and DDoS attack detection. The Atom search algorithm-based hyperparameter optimization and Deep Convolutional Generative Adversarial Network (DCGAN) based Data augmentation is implemented to increase the classifier's efficiency. Our framework makes use of ResNet's superior capacity for learning. From the input data, it automatically extracts the essential elements. This produces better outcomes and avoids the trouble of manually selecting features. The deep network structure of Alexnet leads to a faster training process and avoids the vanishing gradient problem. Therefore, combining these techniques achieves effective DoS and DDoS attack classification results.

In this paper, the primary contributions are listed as follows,

- An effective deep learning technique Alexnet has been proposed to create the intrusion detection system. This model effectively detects various cyber threats like DoS and DDoS attacks.
- To deal with the imbalanced data issue on CCIDS 2019 and UNSW-NB15 Dataset, an effective Deep Convolutional Generative Adversarial Network (DCGAN) is implemented.
- To improve the classification process and achieve high accuracy, the significant features are extracted by the ResNet-50-based technique.
- To decrease the learning complexity of the classifier, the parameters are estimated by the Atom search optimization algorithm, which is used to achieve fast and high-accuracy classification.
- Finally, extensive performance assessments of several existing techniques have been carried out using recall, f1-score, accuracy, and precision metrics.

The remaining portion of this paper is organized as follows: Section 2 provides the relevant studies in this area. Section 3 explains the steps of the suggested strategy. In Section 4, the experimental analysis is entirely detailed. The paper is finally finished with Section 5.

## 2. LITERATURE REVIEW

Convolutional Neural Network (CNN) based hybrid deep learning model was used by Alghazzawi et al. [21] for DDoS attack detection. Initially, they conducted the preprocessing to prepare the raw data for processing. Afterward, the feature selection process based on the chi-squared test was implemented. Then, the features were extracted using CNN; the Bidirectional long/short-term memory (BiLSTM) system was used to detect DDoS attacks. By employing standard performance criteria, including f1-score, recall, precision, and accuracy, the authors evaluated the findings on the CIC-DDoS2019 dataset.

To identify the data representing network traffic patterns, including both regular and DDoS traffic, Aamir and Zaidi [22] devised a clustering-based technique. To extract the features from the dataset, Principal Component Analysis (PCA) was used with two clustering techniques such as k-means and agglomerative techniques. Then a voting technique was implemented to provide the labels for the data to distinguish between assaults and legitimate traffic. After labeling, trained models for future classification are obtained using the supervised machine learning algorithms and Support Vector Machine (SVM), k-Nearest Neighbors (KNN), and Random Forest (RF).

Panigrahi et al. [23] suggested the IDS based on the Consolidated Tree Construction (CTC) method to address the issue of class imbalance. A Supervised Relative Random Sampling (SRRS) technique has been developed for the preprocessing stage to overcome the imbalanced data issue. An Improved Infinite Feature Selection for Multi-class Classification (IIFS-MC) has been implemented to choose the best features of the sample. Finally, J48Consolidated, equipped with CTC, was used to identify potential threats. NSL-KDD, ISCX-IDS2012, and the CICIDS2017 dataset were used for performance assessment.

Khan [24] developed the hybrid deep learning framework to anticipate and categorize harmful cyberattacks in the network using a Convolutional Recurrent Neural Network (CRNN). In this system, CNN conducts convolution to capture local information, while recurrent neural networks (RNNs) capture temporal features to enhance the performance and prediction of the ID system. Experiments were conducted on a publicly accessible CSE-CIC-DS2018 dataset to assess the detection capacity of the suggested approach.

Bi-LSTM with an attention mechanism-based deep learning technique was presented by Fu et al. [25] for traffic anomaly detection. This system initially uses a CNN network to extract sequence information from the traffic data, after which it uses the attention method to reassign the weights of each channel. Finally, Bi-LSTM was used to learn the extracted features. The authors used Adaptive Synthetic Sampling (ADASYN) method to expand minority class samples to address data imbalance issues. Moreover, to decrease the dimensionality of the data, a modified stacked autoencoder was utilized to improve information fusion. For performance assessment, the NSL-KDD dataset was used.

Mighan and Kahani [26] created a hybrid SAE-SVM approach for a cyber-security IDS. The suggested approach employed stacked autoencoder networks for feature extraction and SVM for classification. In the Autoencoder network, multilayer perceptrons were employed for feature extraction. Binary classification and multi-class classification were both done by the authors. The trials were based on accuracy, training time, prediction time, and other performance parameters using the UNB ISCX 2012 IDS dataset.

Based on supervised machine learning, Moualla et al. [27] developed the IDS that contains many Phases. The dataset's unbalanced class problem was first solved by the Synthetic Minority Oversampling Technique (SMOTE) approach. The Extremely Randomized Trees Classifier was then used to choose the key features for each class in the dataset according to the Gini Impurity criterion. Finally, the attacks were classified using the pre-trained extreme learning machine (ELM) model, and the experiments were conducted on the UNSW-NB15 dataset.

## 3. PROPOSED METHODOLOGY

The proposed technique contains four major conceptual components. They are preprocessing, data augmentation, feature extraction, and classification. Initially, the raw data are preprocessed using several techniques. Afterward, Deep Convolutional Generative Adversarial Network (DCGAN) based data augmentation technique is implemented to increase the samples of the minority classes to rectify the imbalanced data issue. Then, the ResNet-50-based deep learning technique is applied to perform the feature extraction. Finally, an optimized Alexnet-based technique is implemented to perform multi-class attack detection. Fig. 1 displays the overall system architecture.

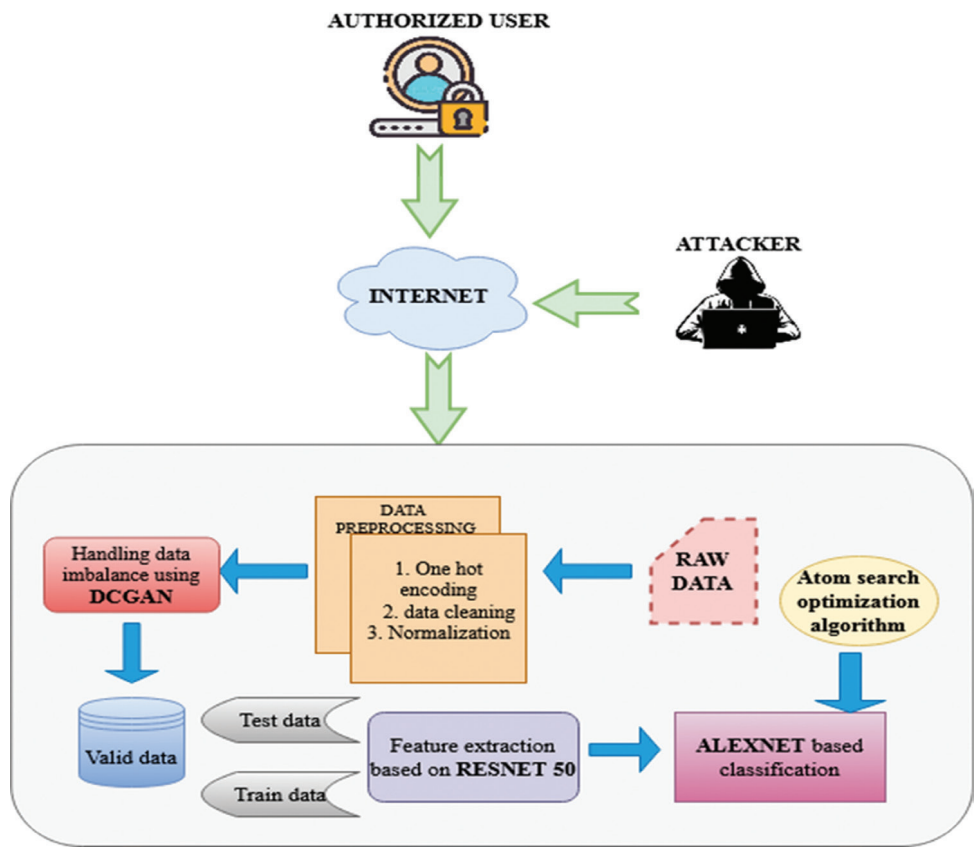

**Fig. 1.** System Framework

### 3.1. PREPROCESSING

Analyzing and cleaning the raw datasets before implementing the proposed technique is necessary because they are typically inaccurate, uniform, or comprehensive.

**Removing socket data**: The CICIDS2019 and UNSW NB15 datasets are in .csv file format, and all socket features—including server and client IP addresses, time stamps, flow ID sources, and destination ports—have been eliminated from the datasets. Because both regular users and intruders have the same IP address; In addition, characteristics like Bwd Bulk Rate (Avg), Bwd Packet/Bulk (Avg), Bwd PSH (Flags), Fwd Packet/Bulk (Avg), Fwd Bytes/Bulk (Avg), Bwd URG, Fwd URG (Flags), and Fwd Bulk Rate (Avg) that have the same value across all rows are eliminated.

**Encoding**: One Hot Encoder is employed in this paper. This encoder creates a new column for each label represented in the dataset and assigns a value of 1 or 0 depending on whether the record falls into that category.

**Normalization**: In this study, Min-Max normalization is used. The primary goal of this normalization is to equalize the values of the features within the range 0 and 1.

$$x' = \frac{x - x_{mn}}{x_{mx} - x_{mn}} \quad (1)$$

Here, $x_{mn}$ and $x_{mx}$ denote minimum and maximum eigenvalue, respectively, the normalized eigenvalue denoted by $x'$ and the original eigenvalue denoted by $x$.

### 3.2. DCGAN-BASED DATA AUGMENTATION

To avoid the imbalance class problem, Deep Convolutional Generative Adversarial Network (DCGAN) is applied for oversampling rare classes. In the UNSW-NB15 dataset, the 'analysis,' 'Shellcode,' 'Worms,' and 'backdoor' classes are increased by the DCGAN technique. Convolutional neural layers are utilized for this network's generator and discriminator models. The following objective function organizes this network:

$$V(D, G) = E_x \sim p_{data(x)}[\log D(x)] E_z \sim p_{z(z)}[\log(1 - D(G(z)))] \quad (2)$$

Here, the actual sample is denoted by $x$, $D(x)$ denotes that the discriminator networks would correctly identify $x$ as an actual sample, $G(z)$ denotes the actual sample from the noise z created by the generator network $G$, and a probability $D(G(z))$ denotes that the discriminator network $D$ will recognize $G(z)$ as an actual sample.

In both the generating and discriminating networks, stride convolution is replaced by the pooling operation in DCGAN. Additionally, global pooling is used in place of the fully connected layer to increase model stability. Following that, Equations (3) and (4) determine the discriminator loss $Ls(D)$ and generator loss $Ls(G)$.

$$Ls(G) = \frac{1}{N} \sum_{i=1}^{N} -\log(D(G(z_i))) \quad (3)$$

$$Ls(D) = \frac{1}{N} \sum_{i=1}^{N} -\log(D(x_i)) - \log(1 - D(G(z_i))) \quad (4)$$

In network training, SGD updates the parameters of discriminator and generator networks based on the loss functions mentioned above. The generator's initial layer is the entire connection layer, followed by the convolution, batch normalization, and leakyReLu activation function, and finally, tanh activates the final layer of convolution. In the discriminator, all the layers follow convolution, batch normalization, and leakyReLu activation, except the first and output layers. Batch normalization is not used in these layers.

### 3.3. RESNET-50-BASED FEATURE EXTRACTION

After the preprocessing, the data is transferred to the ResNet-50 network for feature extraction. The primary goal of this section is to extract meaningful representations from the data to increase the recognition accuracy of the proposed classifier. In this network, the input data is processed through the 1D convolution layers and 16 residual blocks, which are utilized to extract deep features from the data. These blocks are used to address the decomposition and gradient disappearance problems that are typically present in general CNNs. The residual block improves a system's performance without depending on network depth. The following provides the residual function.

$$y = F(x, W) + x \quad (5)$$

Here, the weight is denoted by $W$, the residual input block is denoted by $x$, and the output is denoted by $y$.

The residual block in this network comprises two ReLU activation levels, three 1D convolution layers, and three 1D Batch Normalization layers. Convolution and batch normalization layers match dimensions and skip connections, respectively. Moreover, ReLU layers perform the nonlinear activation, and batch normalization layers are utilized to speed up and stabilize the model. The features are extracted and sent to the average pooling layer in the residual block. The features are pooled, and the pooling results are transferred to the output layer to produce the final features. These features are most intricate and distinctive. To decrease the overfitting, the dropout was set to 0.2.

### 3.4. ALEXNET-BASED INTRUSION CLASSIFICATION

After the feature extraction, the extracted features are transferred to the Alexnet technique for final intrusion classification. The traditional AlexNet contains three fully connected layers, five convolutional layers, three max-pooling layers, and two Local Response Normalization (LRN) layers. To reduce the complication of the network, we used three 1D convolutional layers, two fully connected layers, three pooling layers, and two LRN layers. The convolutional layer is connected to each pooling layer. The random inactivation neuron operation is added to the previous two fully linked layers to prevent the proposed model from overfitting. At last, a softmax layer is the last layer for intrusion classification.

To lower the suggested deep learning framework's computational cost, the pooling layers in the proposed architecture are employed to reduce the size of the feature map. After the first two sessions, response normalization is carried out to lower the test error rate of the suggested network. Network input layers and network input as a whole are both normalized in this layer. The normalization procedure of this layer is explained in the following equation.

$$N_e^x = \frac{b_e^x}{(z + \alpha \sum_{j=\max(0, x-c/2)}^{\min(T-1, x+c/2)} (b_e^x)^2)^\gamma} \quad (6)$$

Here, the normalization of $b_e^x$ neurons activity is denoted by $N_e^x$, calculated at point e using kernel $k$. $z$, $c$, $\alpha$, and $\gamma$ are constants, and the whole kernel's range inside the layer is denoted as $T$. Following that, the output's learned representation is sent into the Softmax layer for multi-class classification aids in calculating classification probabilities. The final classification layer uses these probabilities to categorize the various types of attacks. Then to enhance the classifier's performance, the Atom search optimization algorithm is used to optimize Alexnet's network parameters, including momentum, epoch, initial learning rate, and mini-batch size.

### 3.4.1. Atom Search optimization algorithm

This section implements the combination of the swarm and physics-based algorithm named the Atom Search optimization (ASO) algorithm to optimize the classifier's parameters. The ASO is created using an analysis of the dynamics of molecules and a heuristic algorithm that relies on the kept population. In other words, it may be assumed that the functioning of the suggested ASO relies on the search for the global optima while simulating the mobility of the atoms, which is governed by interactivity and reservation forces. The ASO approach is incredibly straightforward to build and performed exceptionally well. The location of $i^{th}$ atom in a population of $N$ atoms is now computed as,

$$x_i = [x_i^1, \ldots, x_i^d, \ldots x_i^D], \quad i = 1, \ldots \ldots N \quad (7)$$

Here, in a D-dimensional space, the $i^{th}$ atom's $d^{th}$ position component is ($d = 1 \ldots D$). The fitness assessment of the existing population of atoms is used to determine the masses of atoms. That is described as,

$$Ms_i(nt) = e^{-\frac{Fit_i(nt) - Fit_{best}(nt)}{Fit_{worst}(nt) - Fit_{best}(nt)}} \quad (8)$$

$$ms_i(t) = \frac{Ms_i(t)}{\sum_{j=1}^{N} Ms_j(t)} \quad (9)$$

At the nth iteration, all atoms' worst and best fitness is denoted as Fitworst(nt) and Fitbest(nt) correspondingly. An ith atom's and normalized mass are denoted as Msi(nt) and msi(nt), respectively. One way to express the overall force exerted by the other atoms on the $i^{th}$ atom is as follows:

$$F_i^d(nt) = \sum_{j \in Kbest} rand_j F_{ij}^d(nt) \quad (10)$$

$$F_{ij}'(nt) = -\eta(nt)[2(h_{ij}(nt))^{13} - (h_{ij}(nt))^7] \quad (11)$$

Here, the depth function $\eta(t)$ is expressed as,

$$\eta(nt) = \alpha(1 - \frac{nt-1}{mT})^3 e^{-\frac{20nt}{mT}} \quad (12)$$

Here, the maximum number of iterations is denoted by '$mT$,' and the depth weight is denoted as '$a$'. A definition of $h_{ij}$ is expressed as,

$$h_{ij}(nt) = \begin{cases} h_{\min} & \frac{r_{ij}(nt)}{\sigma(nt)} < h_{\min} \\ \frac{r_{ij}(nt)}{\sigma(nt)} & h_{\min} \leq \frac{r_{ij}(nt)}{\sigma(nt)} \leq h_{\max} \\ h_{\max} & \frac{r_{ij}(nt)}{\sigma(nt)} > h_{\max} \end{cases} \quad (13)$$

Here, the upper and lower range of $h$ is denoted as $h_{max}$ and $h_{min}$ correspondingly. The length scale $\sigma(nt)$ is described as

$$\sigma(nt) = \left\| x_{ij}(nt), \frac{\sum_{j \in Kbest} x_{ij}(nt)}{K(nt)} \right\|_2 \quad (14)$$

and

$$\begin{cases} h_{\min} = g_0 + g(nt) \\ h_{\max} = u \end{cases} \quad (15)$$

Here, the drift factor '$df$' has the following definition.

$$d_f(nt) = 0.1 \times \sin(\frac{\pi}{2} \times \frac{nt}{mT}) \quad (16)$$

The definition of the constraint force is,

$$G_i^d(nt) = \lambda(nt)(x_{best}^d(nt) - x_i^d(nt)) \quad (17)$$

The Lagrangian multiplier is described as,

$$\lambda(nt) = \beta e^{-\frac{20nt}{mT}} \quad (18)$$

In this case, the multiplier weight is $\beta$. The $i^{th}$ atom's force at time $nt$ for each repetition is expressed as,

$$F_r = F_i + G_i \quad (19)$$

$$a_i^d(nt) = \frac{F_i^d(nt)}{m_i(nt)} = -\alpha(1 - \frac{nt-1}{mT})^3 e^{-\frac{20nt}{mT}}$$

$$\times \sum_{j \in Kbest} \frac{rand_j \left[2 \times (h_{ij}(nt))^{13} - (h_{ij})^7\right]}{m_i(nt)} \frac{(x_i^d(nt) - x_i^d(nt))}{\|x_i(nt), x_j(nt)\|2} \quad (20)$$

$$+ \beta e^{-\frac{20nt}{mT}} \frac{x_{best}^d(nt) - x_i^d(nt)}{m_i(nt)}$$

In this case, the multiplier weight is denoted as *β*, and the depth weight is denoted as '*a*'. The first *K* atoms with the highest fitness values make up the subset of an atom population known as $K_{best}$; *K* is denoted by

$$K(nt) = N - (N-2) \times \sqrt{\frac{nt}{mT}} \quad (21)$$

The $i^{th}$ atom's speed and position at time $nt + 1$ are updated accordingly.

$$v_i^d(nt+1) = rand_i^d v_i^d(nt) + a_i^d(nt) \quad (22)$$

$$x_i^d(nt+1) = x_i^d(nt) + v_i^d(nt+1) \quad (23)$$

Here, the atom's velocity is denoted by $v_i^d$, the acceleration of the atom is denoted by $a_i^d(nt)$, and $i^{th}$ atom's position is denoted by $x_i^d(nt)$. Finally, the algorithm's best optimal value initializes the classifier's hyperparameters. The optimized values are momentum=0.9, weight decay=0.005, epoch=100, initial learning rate=0.001, and mini-batch size=32.()

## 4. RESULTS AND DISCUSSION

The performance and efficiency of the proposed IDS are assessed by several experiments on the UNSW-NB15 and CICIDS2019 datasets using typical performance metrics, which are discussed in this section. The experiments were carried out with Python programming language, and Keras was used to run all of the simulations with Tensorflow as the backend, using an Intel Core i7-7700 CPU and 32 GB RAM.

### 4.1. DATASET DESCRIPTION

#### 4.1.1. UNSW-NB15dataset

The Australian Centre for Cyber Security research team developed the UNSW-NB15 dataset. More than 2.5 million network packets are used to replicate this data set. This data set includes non-anomalous packets and nine other types of attacks (Exploits, Reconnaissance, DoS, Generic, Shellcode, Fuzzers, Backdoors, Worms, and Analysis). The data set is highly unbalanced because more than 87% of the packets are non-anomalous.

#### 4.1.2. CICIDS2019 dataset

This dataset includes a variety of DDoS assaults that can be conducted via TCP/UDP application layer protocols. The taxonomy of attacks in the dataset is carried out in terms of exploitation-based and reflection-based assaults. More than 80 flow features are included in the dataset. The dataset was gathered over two various days for testing and training analysis. The assaults in the dataset include DDoS attacks using DNS, NTP, NetBIOS, SYN, MSSQL, UDP-Lag, LDAP, and SNMP.

### 4.2. PERFORMANCE METRICS

The proposed intrusion model's performance was assessed utilizing its accuracy, precision, recall, and F-score criteria. An overview of the classification metrics is given in this section. The proportion of accurately classified data out of all classified data is how accurately something is classified. The accuracy of an optimistic prediction is estimated by precision. A low rate of false positives indicates high precision. Recall counts the instances that are classified correctly as positive. F-score integrates recall and precision. It is possible to define it as the average of precision and recall. The measures mentioned above can be expressed mathematically as,

$$Accuracy = \frac{TruPsv + TruNeg}{TruPsv + TruNeg + FlsPsv + FlsNeg} \quad (24)$$

$$\Pr ecision = \frac{TruPsv}{TruPsv + FlsPsv} \quad (25)$$

$$\operatorname{Re} call = \frac{TruPsv}{TruPsv + FlsNeg} \quad (26)$$

$$F1 - Score = \frac{2 * precision * recall}{precision + recall} \quad (27)$$

Here, true positives (TruPsv) are the class or value of occurrence that was accurately predicted. False positives (FlsPsv) occur when the actual class differs from the anticipated one, and false negatives (FlsNeg) are values for no events that were mistakenly predicted to occur. Correctly estimated no event values are referred to as true negative (TruNeg) values.

### 4.3. TRAINING AND TESTING

In this method, the entire dataset is split into two groups: one is used to train the network (70% of data), and the other is used to test it (30%). The training set is used to train the proposed model for 100 epochs to decrease the error in the model to the fullest degree possible, and a 0.001 learning rate allows the model to train faster. Moreover, the proposed ASO algorithm optimizes momentum=0.9, weight decay=0.005, mini-batch size=32, and the bias of each layer in the classifier. The training and testing accuracy and loss of the proposed approach for both datasets are shown in Figs. 2 and 3

The testing and training accuracy of the presented technique for both datasets are shown in Figs. 2 and 3, and the loss values range from 0.001 to 0.004. When evaluated using benchmark datasets, the suggested model's performance displays a similar pattern, demonstrating the model's ability to predict attacks from other categories besides those stated. Additionally, as the image shows, the training loss is relatively high initially but gradually decreases as the training progresses. Once the training epoch reaches 20, the error usually decreases more slowly.

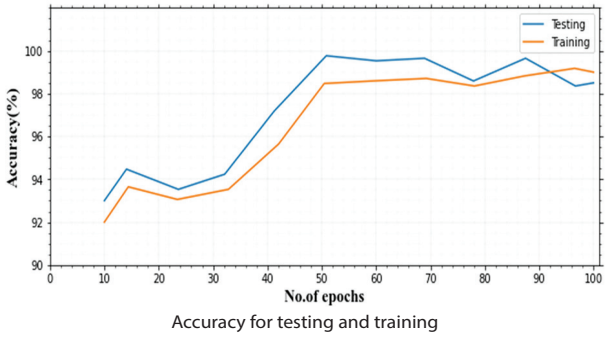
Accuracy for testing and training

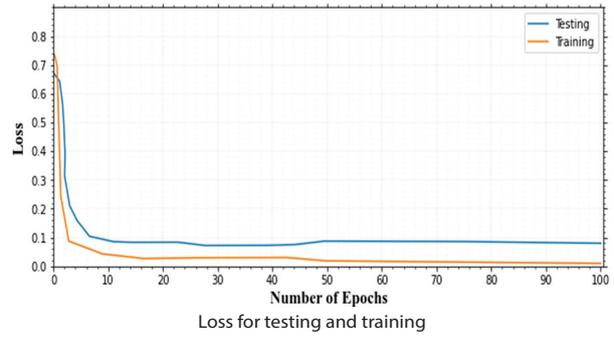
Loss for testing and training

**Fig. 2**. (a) Testing and training accuracy, (b) testing and training loss for the CICIDS2019 dataset

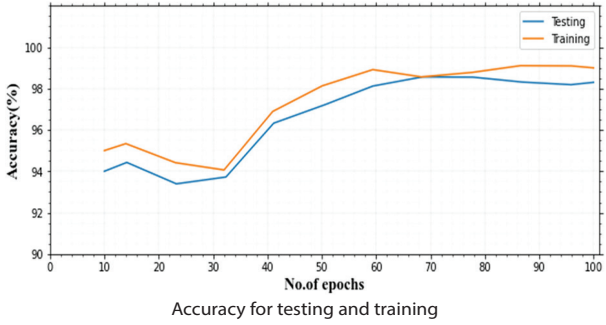
Accuracy for testing and training

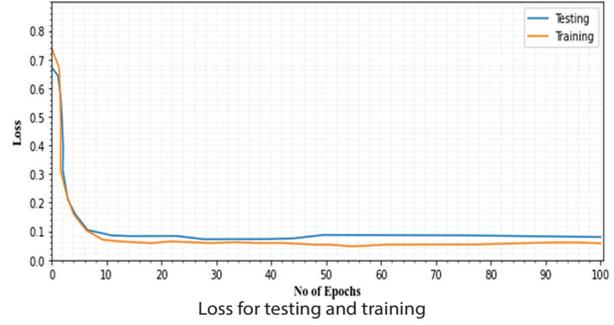
Loss for testing and training

**Fig. 3**. (a) Testing and training accuracy, (b) Testing and training loss for UNSW-NB15 dataset

### 4.4. PERFORMANCE EVALUATION ON THE CICIDS2019 DATASET

Various tests have been performed on the CICIDS2019 dataset to assess the effectiveness of the suggested approach. The multi-class classification result of the proposed approach is given in Table 1.

**Table 1.** Multi-class classification of the proposed approach on the CICIDS2019 dataset

| Attack types | F1-score | Recall | Precision | Accuracy |
|---|---|---|---|---|
| Normal | 99.88 | 99.89 | 99.87 | 99.92 |
| DNS | 99.25 | 99.27 | 99.24 | 99.29 |
| NTP | 99.33 | 99.4 | 99.38 | 99.41 |
| NetBIOS | 99.75 | 99.81 | 99.78 | 99.89 |
| SYN | 99.76 | 99.77 | 99.74 | 99.78 |
| MSSQL | 98.72 | 98.77 | 98.77 | 98.79 |
| UDP | 99 | 99.04 | 99 | 99.09 |
| LDAP | 99.18 | 99.24 | 99.2 | 99.25 |
| SNMP | 98.98 | 98.92 | 98.95 | 98.98 |
| UDP-LAG | 98.97 | 98.94 | 98.96 | 98.99 |

Table 1 shows that the proposed approach attains superior outcomes for all attacks on the CICIDS2019 dataset regarding f1-score, recall, precision, and accuracy. Particularly, Normal, NetBIOS, and SYN classes attain superior results with correspondingly 99.92%, 99.89%, and 99.78% accuracy. Moreover, the classification performance on UDP and LDAP provides the best performance. MSSQL detection performance is average compared to all the classes, with 98.79% accuracy, 98.77% precision, 98.77% recall, and 98.72% f1-score.

A graphical representation of Table 1 is shown in Fig. 4.

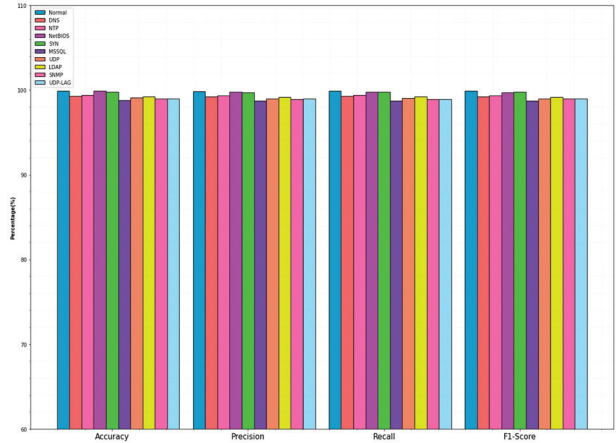

**Fig. 4.** Multi-class classification of the CICIDS2019 dataset

**Table 2.** Comparison of the proposed approach on the CICIDS2019 dataset

| Technique | F1-score | Recall | Precision | Accuracy |
|---|---|---|---|---|
| Adaboost Regression [28] | - | 96.74 | 95.93 | 95.87 |
| XNN[29] | 99 | 99.2 | 99 | 99.3 |
| LSTM [30] | 97.8 | 98 | 98.1 | 98 |
| KNN [31] | 97 | 97 | 96 | 98 |
| Proposed | 99.28 | 99.30 | 99.28 | 99.33 |

Table 1 shows the differentiation of the performance of the suggested work with other standard techniques tested under the CICIDS2019 dataset. The table shows that the IDS model based on the proposed approach incurs the best results in terms of recall, accuracy and precision, and f1-score. Compared to all other methods, the performance of the Adaboost technique is poor (95.87% accuracy), and the XNN technique provides the best performance with 99% f1-score, 99.2% recall, 99% precision, and 99.3% accuracy. However, these are just as good as our proposed approach. The Long short-term memory (LSTM) and K-nearest Neighbor (KNN) perform similarly with 98% accuracy. A graphical representation of Table 1 is shown in Fig. 5.

better values for all the attack classes. All the classes attain above 99% accuracy for all the classes. Specifically, the proposed approach classifies normal, Analysis, and Shellcode with 99.89%, 99.81%, and 99.79% accuracy. Compared to all attacks, the classification performance on the proposed approach of Reconnaissance and Exploits is average, with 99% accuracy. These values are the best. However, compared to all other classes, these values are shallow. The graphical representation of this table is presented in Fig. 6.

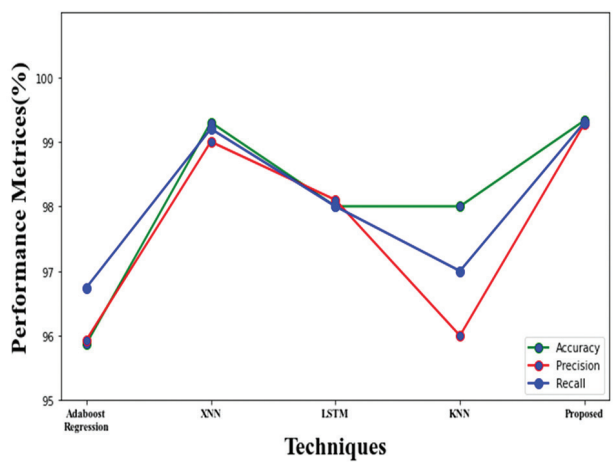

**Fig. 5.** Comparison of the proposed approach on the CICIDS2019 dataset

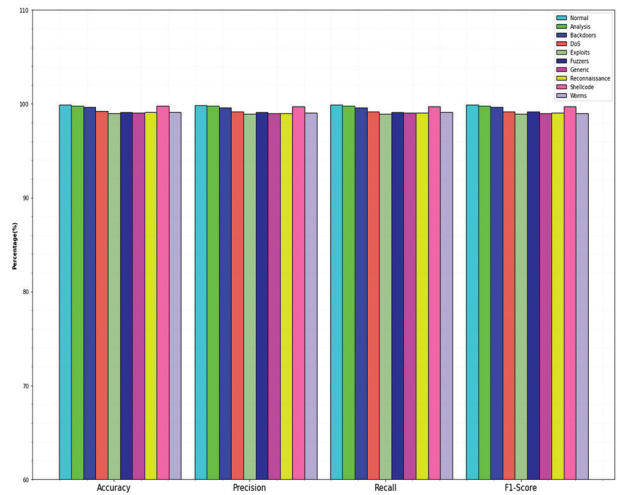

**Fig. 6.** Multi-class classification of the UNSW-NB15 dataset

### 4.5. PERFORMANCE EVALUATION ON UNSW-NB15 DATASET

On the UNSW-NB15 dataset, numerous experiments are performed to analyze the effectiveness of the presented strategy. The multi-class classification result of the proposed approach is given in Table 3.

**Table 3.** Multi-class classification of the proposed approach on the UNSW-NB15 dataset

| Attack types | F1-score | Recall | Precision | Accuracy |
|---|---|---|---|---|
| Normal | 99.88 | 99.88 | 99.87 | 99.89 |
| Analysis | 99.79 | 99.76 | 99.78 | 99.81 |
| Backdoors | 99.64 | 99.63 | 99.61 | 99.69 |
| DoS | 99.18 | 99.19 | 99.16 | 99.21 |
| Exploits | 98.93 | 98.94 | 98.91 | 99 |
| Fuzzers | 99.15 | 99.11 | 99.08 | 99.12 |
| Generic | 99 | 99.02 | 99 | 99.04 |
| Reconnaissance | 99.04 | 99.06 | 99 | 99.1 |
| Shellcode | 99.73 | 99.74 | 99.71 | 99.79 |
| Worms | 99 | 99.09 | 99.04 | 99.12 |

Table 3 shows that the proposed approach's multi-classification performance is superior and achieves

**Table 4.** Comparison of the proposed approach on the UNSW-NB15 dataset

| Technique | F1-score | Recall | Precision | Accuracy |
|---|---|---|---|---|
| SVM-ANN [32] | 87.01 | - | 96.89 | 97.98 |
| RLF-CNN [33] | - | 89.3 | - | 88.7 |
| ELM [34] | 96.08 | - | - | 98.19 |
| ANN[35] | - | - | - | 97.89 |
| Proposed | 99.33 | 99.34 | 99.31 | 99.37 |

According to the residual blocks, a deep learning algorithm allows the construction of deeper networks to identify more critical network traffic characteristics. Our model outperforms existing deep learning techniques, as seen in Table 5 and Fig. 6. CNNs can perform better in network intrusion detection with residual learning. Due to issues with a class imbalance in the training set, none of the other models perform well in terms of minor classes. To overcome the abovementioned problem, our model uses a class imbalance strategy based on DCGAN. In all other existing techniques, the performance of RLF-CNN is mildly decreased due to its decreased weights in the loss function. Moreover, the performance of ANN (Artificial Neural Network) and SVM-ANN (Support vector machine-ANN) are similar. However, it is not more than the proposed approach. A graphical representation of the accuracy comparison is displayed in Fig. 7.

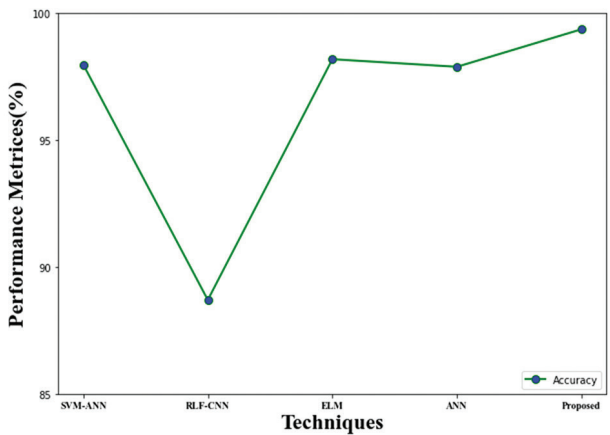

**Fig. 7.** Accuracy comparison of the proposed approach on the UNSW-NB15 dataset

### 4.6. COMPARISON OF DATA AUGMENTATION

The impact of data augmentation on the classification criteria for both datasets is illustrated in Fig. 8. The data augmentation technique enhances the classifier performance, as is shown in this graph. Using a DC-GAN-based data augmentation strategy, the classifier in the CICIDS2019 dataset achieves 99.33% accuracy. It only accomplishes 99.12% without DCGAN. Using data augmentation approaches, the classifier for the UNSW-NB15 dataset achieves 99.37% accuracy. Compared to training samples without data augmentation, the number of training samples produced by this method is much higher.

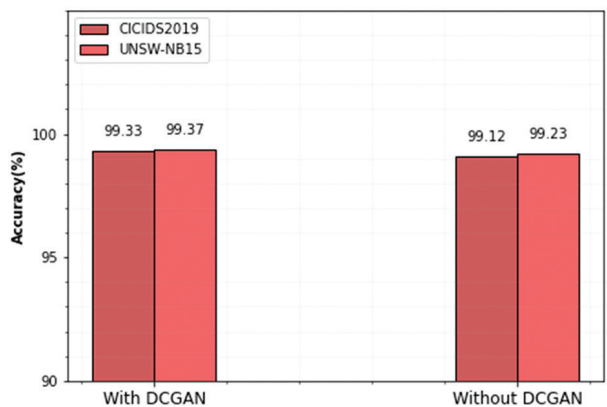

**Fig. 8.** Analysis of Data Augmentation Technique

In the absence of data augmentation, the same group of training instances is utilized for each epoch; however, when data augmentation is present, different training instances are generated for each epoch. Because of this, the proposed algorithm employs the DCGAN data augmentation technique to perform better and attain higher accuracy in both datasets.

### 5. CONCLUSION

This research presented the Deep learning-based intrusion detection system, which utilized a Resnet-50-based effective technique to extract features from the network data. The suggested IDS is validated using the UNSW-NB15 and CICIDS2019 datasets—a deep learning system built on the optimized Alexnet to identify the attacks effectively. The UNSW-NB15 and CICIDS2019 datasets had the highest accuracy, each at 99.37%. The suggested model performs admirably in a multi-class setting regarding f1-score, recall, precision, and accuracy measures. The research conducted here also aimed to offer guidance in selecting the optimum dataset for the model. The optimal dataset for the suggested model has been determined to be the UNSW-NB15 dataset. In the future, more assessment metrics will be conducted to analyze the system's efficacy with less time and resource usage.

**Conflict of interests**: The authors declare that they have no known competing financial interests or personal relationships that could have appeared to influence the work reported in this paper.

**Acknowledgments**

We declare that this manuscript is original, has not been published before, and is not currently being considered for publication elsewhere.

**Availability of data and material**: Not applicable

**Code availability**: Not applicable

**Authors' contributions**: The author confirms sole responsibility for the following: study conception and design, data collection, analysis and interpretation of results, and manuscript preparation.

**Ethics approval**: This material is the author's original work, which has yet to be previously published elsewhere. The paper reflects the author's research and analysis truthfully and completely.